\documentclass[11pt,a4paper]{article}
\usepackage[hyperref]{emnlp2020}
\usepackage{times}
\usepackage{latexsym}
\usepackage{multirow}
\usepackage{subfigure}
\usepackage{graphicx}
\usepackage{lscape}
\usepackage{marvosym}
\usepackage{booktabs}
\usepackage{amsmath}
\usepackage{color,soul}

\usepackage{microtype}

\aclfinalcopy

\title{BERT-QE: Contextualized Query Expansion for Document Re-ranking}

\author{Zhi Zheng$^{1,3}$, Kai Hui$^2$\thanks{\quad This work has been done before joining Amazon.}, Ben He$^{1,3}$\textsuperscript{\Letter}, Xianpei Han$^3$, Le Sun$^3$\textsuperscript{\Letter}, Andrew Yates$^4$\\
	$^{1}$ University of Chinese Academy of Sciences, Beijing, China \\
	$^{2}$ Amazon Alexa, Berlin, Germany\\ 
	$^{3}$ Institute of Software, Chinese Academy of Sciences, Beijing, China\\
    $^{4}$ Max Planck Institute for Informatics, Saarbr\"ucken, Germany \\
\small {\tt zhengzhi18@mails.ucas.ac.cn}, \hspace{0.15cm}  {\tt kaihuibj@amazon.com }  \\ 
\small  {\tt benhe@ucas.ac.cn}, \hspace{0.15cm} {\tt \{xianpei, sunle\}@iscas.ac.cn}, \hspace{0.15cm} {\tt ayates@mpi-inf.mpg.de}
}

\date{}

\begin{document}
\maketitle
\begin{abstract}

Query expansion aims to mitigate 
the mismatch between
the language used in a query and 
in a document. However, query expansion methods can suffer from introducing 
non-relevant information when expanding 
the query.
To bridge this gap, inspired by recent advances in applying contextualized models like BERT to the document retrieval task, this paper proposes a novel query expansion model that leverages the strength of 
the BERT model to select
relevant document chunks for expansion.
In evaluation on the standard TREC Robust04 and GOV2 test collections,
the proposed BERT-QE model significantly outperforms BERT-Large models.
\end{abstract}

\section{Introduction}\label{sec:intro}

In information retrieval, 
the language used in a query and in a document
differs 
in terms of verbosity,  
formality, and even the 
format (e.g., the use of keywords in a query
versus the use of natural language in an article from Wikipedia).
In order to reduce this gap, different query expansion methods have been proposed and have enjoyed success in improving document rankings.
Such methods commonly take a pseudo relevance feedback (PRF) approach in which the query is expanded using top-ranked documents
and then the expanded query is used to rank the search results~\cite{rocchio1971relevance,DBLP:conf/sigir/LavrenkoC01,DBLP:phd/ethos/Amati03,DBLP:conf/sigir/MetzlerC07} .

Due to their reliance on pseudo relevance information,
such expansion methods suffer from any non-relevant information in the feedback documents, which
could pollute the query after expansion.
Thus, selecting and re-weighting the 
information pieces from PRF according to their relevance 
before re-ranking are crucial for the effectiveness of the query expansions. 
Existing works identify expansion tokens according to the language model on top of feedback documents, as in RM3 ~\cite{DBLP:conf/sigir/LavrenkoC01}, extract the topical terms from feedback documents that diverge most
from the corpus language model~\cite{DBLP:phd/ethos/Amati03}, or extract concepts for expansion~\cite{DBLP:conf/sigir/MetzlerC07}.
In the context of neural approaches, the recent neural PRF architecture~\cite{DBLP:conf/emnlp/LiSHWHYSX18} uses feedback documents directly for expansion.
All these methods, however, are under-equipped to accurately evaluate
the relevance of information pieces used for expansion.
This can be caused by the mixing
of relevant and non-relevant information in
the expansion, like the tokens in RM3~\cite{DBLP:conf/sigir/LavrenkoC01} and the
documents in NPRF~\cite{DBLP:conf/emnlp/LiSHWHYSX18}; or by the
facts that the models used for selecting and re-weighting the expansion information
are not powerful enough, as they are essentially scalars based on counting.

Inspired by the recent advances of 
pre-trained contextualized models like BERT on the ranking task~\cite{DBLP:conf/emnlp/YilmazYZL19, DBLP:journals/corr/abs-2003-06713}, 
this work attempts to develop query expansion models based on BERT with the goal of more effectively using the relevant information from PRF.
In addition, 
as indicated in previous studies~\cite{DBLP:journals/corr/abs-1904-07531,DBLP:conf/sigir/DaiC19}, the
(pre-)trained BERT-based ranking models have a strong ability to identify highly relevant chunks within documents. 
This actually provides advantages in choosing text chunks for expansion by
providing more flexibility in terms of the granularity for expansions,
as compared with using tokens~\cite{DBLP:conf/sigir/LavrenkoC01}, concepts with one or two words~\cite{DBLP:conf/sigir/MetzlerC07},
or documents~\cite{DBLP:conf/emnlp/LiSHWHYSX18}.

Given a query and a list of feedback documents from an initial ranking (e.g., from BM25), we propose to re-rank the documents in
three sequential phases.
In phase one, the documents are re-ranked with a fine-tuned BERT model and the
top-ranked documents are used as PRF documents;
in phase two, these PRF documents are decomposed into text chunks with fixed length (e.g., 10), and the relevance of individual chunks are evaluated;
finally, to assess the relevance of a given document, the selected chunks and original query are used to score the document together. 
To this end, a novel query expansion model, coined as BERT-QE, based on 
the contextualized model is developed.

Contributions of this work are threefold.
1) A novel query expansion model is proposed
to exploit the strength of contextualized model BERT
in identifying relevant information from feedback documents;
2) Evaluation on two standard TREC test collections, namely, Robust04 and GOV2, demonstrates 
that the proposed BERT-QE-LLL could advance the performance of BERT-Large significantly on both shallow and deep pool, when using BERT-Large in all three phases;
3) We further trade-off the efficiency and effectiveness, 
by replacing BERT-Large with smaller BERT architectures and
demonstrate that, with a smaller variant of BERT-QE, e.g., BERT-QE-LMT, one
could outperform BERT-Large significantly on shallow pool with as least as an extra 3\% computational cost;
meanwhile, a larger variant, e.g., BERT-QE-LLS,
could significantly outperform BERT-Large on both shallow and deep pool
with 30\% more computations.

\begin{figure*}[!t]
    \centering
    \includegraphics[width=0.6\textwidth]{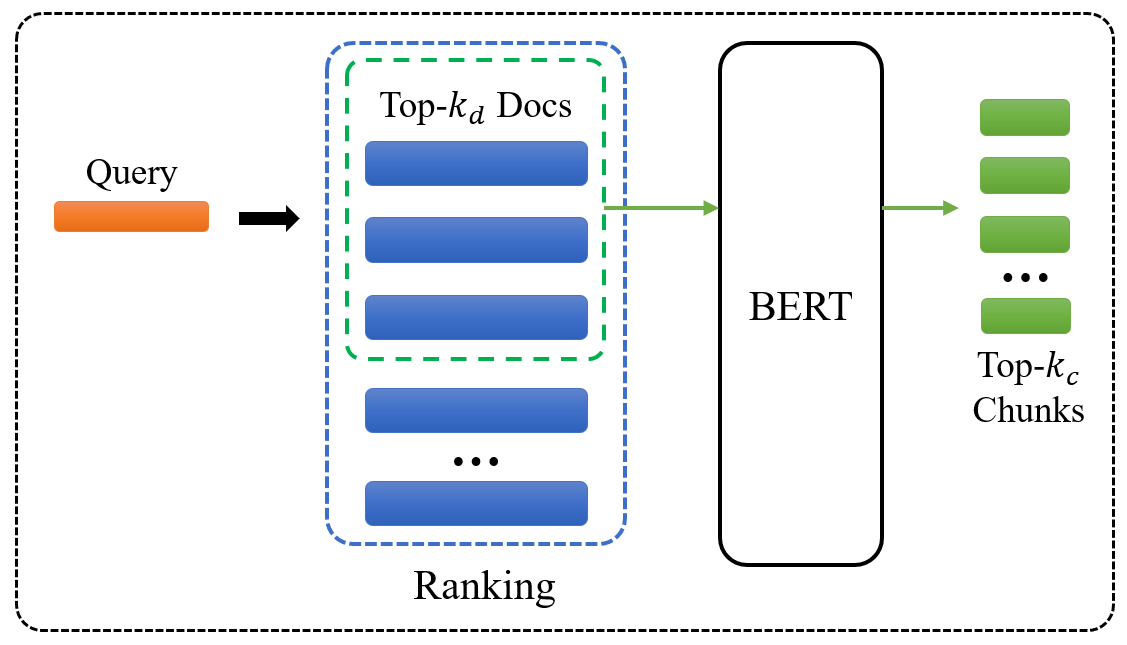}
    \caption{Chunk selection for query expansion in phase two.}
    \label{fig.phase2}
\end{figure*}

\section{Method}\label{sec.method}
In this section we describe BERT-QE,
which takes a ranked list of documents as input (e.g., from an unsupervised ranking model) and outputs a re-ranked list based on the expanded query.

\subsection{Overview}\label{sec.overview}
There are three phases in the proposed BERT-QE. Namely, \textbf{phase one}: the first-round re-ranking of the 
documents using a BERT model; \textbf{phase two}: chunk selection for query expansion from the top-ranked documents; and \textbf{phase three}: the final re-ranking using the selected expansion chunks.
The essential parts of the proposed BERT-QE are the second and third phases, which are introduced in detail in Sections~\ref{sec.chunkselect} and~\ref{sec.finalrerank}.
Without loss of generality,
a fine-tuned BERT model serves as the backbone of the proposed BERT-QE model and is used in all three phrases. 
We describe the fine-tuning process and phase one before describing phases two and three in more detail.

\noindent\textbf{Fine-tuning BERT model.}
Similar to~\cite{DBLP:conf/emnlp/YilmazYZL19},
a BERT model (e.g., BERT-large)
is first initialized using a checkpoint 
that has been trained on MS MARCO~\cite{bajaj2018ms}.
The model is subsequently fine-tuned
on a target dataset (e.g., Robust04).
This choice is to enable comparison with
the best-performing BERT model, such as a fine-tuned BERT-Large~\cite{DBLP:conf/emnlp/YilmazYZL19}.
Before fine-tuning the BERT model on a target dataset, we first use the aforementioned model
trained on MS MARCO
to identify the top-ranked passages in this dataset.
These selected query-passage pairs are then used to fine-tune BERT using the loss function as in Equation~(\ref{eq.pointwiseloss}).

  \begin{equation}\label{eq.pointwiseloss}
      \mathcal{L} = - \sum_{i\in I_{pos}}{\log(p_i)-\sum_{i\in I_{neg}}{\log(1-p_i)}}
 \end{equation}
 Therein, $I_{pos}$ and $I_{neg}$ are sets of indexes of the relevant and non-relevant documents, respectively, and $p_i$ is the probability of the document $d_i$ being relevant to the query.
This configuration is similar to ~\citet{DBLP:conf/sigir/DaiC19}, 
with the difference that we use
only passages with the highest scores 
instead of all passages.
In our pilot experiments, this leads to comparable effectiveness but with a shorter training time.

\noindent\textbf{Phase one.}
Using the fine-tuned BERT model, we re-rank a list of documents 
from an unsupervised ranking model for use in the second phase.
As shown in Equation~(\ref{eq.firstround}),
given a query $q$ and a document $d$, 
$\mathit{rel}(q,d)$ assigns 
$d$ a relevance score by modeling the
concatenation of the query and the document using the fine-tuned
BERT.
 The ranked list is obtained by ranking the documents with respect to these relevance scores.
We refer the reader to prior works describing BERT and ranking with BERT for further details~\cite{DBLP:conf/naacl/DevlinCLT19,DBLP:journals/corr/abs-1901-04085}.

\begin{equation}\label{eq.firstround}
\mathit{rel}(q,d) = \operatorname{BERT}(q,d)
\end{equation}

\subsection{Selecting Chunks for Query Expansion}~\label{sec.chunkselect}
In the \textbf{second phase}, the top-$k_d$ documents from the first phase are employed as feedback documents and $k_c$ chunks of relevant text are extracted from them.
This phase is illustrated in Figure \ref{fig.phase2}.
In more detail,
a sliding window spanning $m$ words is used to decompose each feedback document into overlapping chunks where
two neighboring chunks are overlapped by up to $m/2$ words.
The $i$-th chunk is denoted as $c_i$. 
As expected, these chunks are a mixture of relevant and non-relevant text pieces due to the lack of supervision signals. 
Therefore, the fine-tuned BERT model from Section~\ref{sec.overview} is used to
score each individual chunk $c_i$, as indicated in Equation~(\ref{eq.chunkscore}).
The top-$k_c$ chunks with the highest scores are selected.
These $k_c$ chunks, which are the output from phase two, serve as a distillation of the feedback information in the feedback documents from phase one.
We denote the chunks as $\mathcal{C}=[c_0, c_1, \cdots, c_{k_{c}-1}]$.

\begin{equation}\label{eq.chunkscore}
    \mathit{rel}(q,c_i) = \operatorname{BERT}(q,c_i)
\end{equation}

\subsection{Final Re-ranking using Selected Chunks}~\label{sec.finalrerank}
\begin{figure*}[!t]
    \centering
    \includegraphics[width=0.6\textwidth]{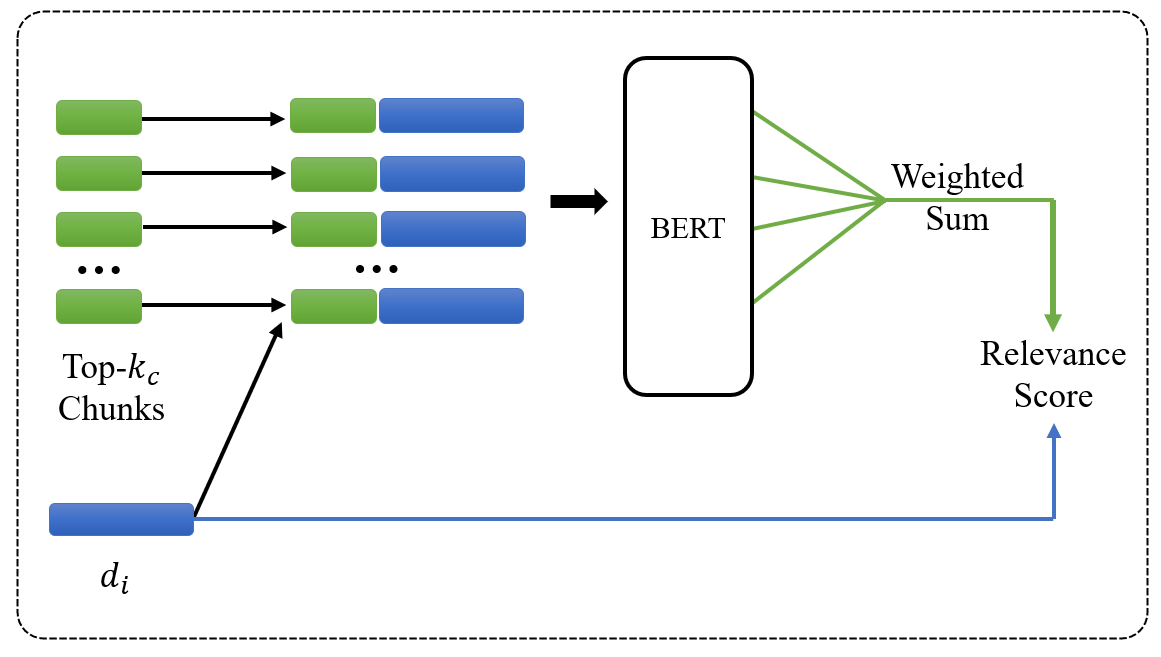}
    \caption{Re-rank documents using selected chunks in phase three.}
    \label{fig:term_chunk_figure}
\end{figure*}

In \textbf{phase three}, the chunks selected from phase two are used in combination with the original query to compute a final re-ranking.
This process is illustrated in Figure~\ref{fig:term_chunk_figure}.

\noindent\textbf{Evaluating the relevance of a document using the selected feedback chunks.}
For each individual document $d$, 
the $k_c$ chunks selected in phase two
are used to assess its relevance separately, and the $k_c$
evaluations are thereafter aggregated to generate the document's relevance score.
As described in Equation~(\ref{eq.feedbackscore}),
the fine-tuned BERT model from Section~\ref{sec.overview} is used to compute 
$\mathit{rel}(c_i, d)$,
which are further aggregated into a relevance score $\mathit{rel}(\mathcal{C},d)$.
Akin to~\cite{DBLP:conf/emnlp/LiSHWHYSX18}, the relevance of individual chunks are incorporated as weights by using the softmax function $\operatorname{softmax}_{c_i\in \mathcal{C}}(.)$ among all chunks in $\mathcal{C}$
on top of the $\mathit{rel}(q,c_i)$.
\begin{equation}\label{eq.feedbackscore}
    \mathit{rel}(\mathcal{C},d) = \sum_{c_i \in \mathcal{C}}\operatorname{softmax}_{c_i\in \mathcal{C}}(\mathit{rel}(q,c_i)) \cdot \mathit{rel}(c_i, d)
\end{equation}

\noindent\textbf{Combining $\mathit{rel}(\mathcal{C},d)$ with $\mathit{rel}(q,d)$.}
To generate the ultimate relevance score $\mathit{rel}(q,\mathcal{C},d)$ for $d$,
akin to the established PRF models like RM3~\cite{DBLP:conf/sigir/LavrenkoC01} and NPRF~\cite{DBLP:conf/emnlp/LiSHWHYSX18},
the relevance scores based on the feedback and the original query are combined as in Equation~(\ref{eq.combine}).
$\alpha$ is a hyper-parameter, governing the relative importance of the two parts.

\begin{equation}\label{eq.combine}
    \mathit{rel}(q,\mathcal{C},d) = (1 -\alpha) \cdot \mathit{rel(q,d)}+ \alpha \cdot\mathit{rel}(\mathcal{C},d)  
\end{equation}

We note that the same fine-tuned BERT model does not necessarily need to be used in each phase.
In our experiments, we consider the impact of using different BERT variants from Table~\ref{tab.bertconfigure} in each phase.
For example, phases one and three might use the BERT-Large variant, while phase two uses the BERT-Small variant with fewer parameters.

\section{Experimental Setup}\label{sec.exp_setting}
In this section, we describe our experiment configurations. Source code, data partitions for cross-validation, result files of initial rankings, and the trained models are available online\footnote{\url{https://github.com/zh-zheng/BERT-QE}}.

\subsection{Dataset and Metrics}

Akin to~\cite{DBLP:conf/cikm/GuoFAC16,DBLP:conf/emnlp/YilmazYZL19}, we use the standard Robust04~\cite{DBLP:conf/trec/Voorhees04b} and GOV2~\cite{DBLP:conf/trec/ClarkeCS04} test collections.
Robust04 consists of 528,155 documents and GOV2 consists of 25,205,179 documents. 
We employ 249 TREC keyword queries for Robust04 and 150 keyword queries for GOV2.
Akin to~\cite{DBLP:conf/emnlp/YilmazYZL19},
in this work, 
all the rankings from BERT-based models, including
the proposed models and the baselines,
have been interpolated 
 with the initial ranking scores (DPH+KL in this work) in the same way wherein the hyper-parameters are tuned in cross-validation\footnote{The details of the interpolation for BERT-QE are included in Appendix.}.
We report P@20, NDCG@20 to enable the comparisons on the shallow pool; and MAP@100, MAP@1000 are reported for deep pool.
In addition, statistical significance for paired two-tailed t-test is reported, where the superscripts $***$, $**$ and $*$ denote the significant level at 0.01, 0.05, and 0.1, respectively.

\subsection{Initial Ranking} 

\noindent\textbf{DPH+KL} is used as the ranking model to generate the initial ranking. DPH is an unsupervised retrieval model~\cite{DBLP:conf/trec/AmatiABGG07} 
derived from the divergence-from-randomness framework. 
DPH+KL ranks the documents with DPH after expanding the original queries with Rocchio's query expansion using Kullback-Leibler divergence~\cite{DBLP:phd/ethos/Amati03,rocchio1971relevance}, as implemented in the Terrier toolkit~\cite{macdonald2012puppy}. 
Its results are also listed for comparison.

\subsection{Models in Comparisons}\label{sec.models_comparison}

\noindent\textbf{Unsupervised query expansion models,}
like Rocchio's query expansion~\cite{rocchio1971relevance} with the KL divergence model \cite{DBLP:phd/ethos/Amati03}, and RM3~\cite{DBLP:conf/sigir/LavrenkoC01}, are employed as a group of baseline models,
wherein the query is expanded by selecting terms from top-ranked documents from the initial ranking.

- \textbf{BM25+RM3} is also used as a baseline model, which follows the experimental settings from~\cite{DBLP:conf/emnlp/YilmazYZL19}, and the implementation from
Anserini~\cite{DBLP:conf/ecir/LinCTCCFIMV16} with default settings is used.

- \textbf{QL+RM3} is the query likelihood language model with RM3 for PRF~\cite{DBLP:conf/sigir/LavrenkoC01}, for which the Anserini's~\cite{DBLP:conf/ecir/LinCTCCFIMV16} implementation with default settings is used.

\noindent\textbf{Neural ranking models.} We also include  different neural ranking models for comparisons. 

- \textbf{SNRM}~\cite{DBLP:conf/cikm/ZamaniDCLK18} is a standalone neural ranking model by introducing a sparsity property to learn a latent sparse representation for each query and document. The best-performing version of SNRM with PRF is included for comparison.

- \textbf{NPRF}~\cite{DBLP:conf/emnlp/LiSHWHYSX18} is an end-to-end neural PRF framework that can be used with existing neural IR models, such as DRMM~\cite{DBLP:conf/cikm/GuoFAC16}. The best-performing variant NPRF$_{ds}$-DRMM is included for comparison.

- \textbf{CEDR}~\cite{DBLP:conf/sigir/MacAvaneyYCG19} incorporates the classification vector of BERT into existing neural models. The best-performing variant CEDR-KNRM is included for comparison.

- \textbf{Birch}~\cite{DBLP:conf/emnlp/YilmazYZL19} is a re-ranking approach by fine-tuning BERT successively on the MS MARCO and MicroBlog (MB) datasets. 
The best-performing version 3S: BERT(MS MARCO$\rightarrow$MB), denoted as Birch(MS$\rightarrow$MB) for brevity, is included for comparison.

- \textbf{BERT-Large and BERT-Base} in the MaxP configuration are fine-tuned on the training sets with cross-validation as described in Section~\ref{sec.overview}.

\subsection{Variants of BERT}\label{sec:bert_sizes}
Different variants of BERT models with different configurations are employed.
We list the key hyper-parameters of each variant in Table~\ref{tab.bertconfigure}, namely, the number of hidden layers, the hidden embedding size, and the number of attention heads, which are denoted as $L$, $H$ and $A$, respectively\footnote{Note that the BERT-Small corresponds to BERT-Mini in \url{https://github.com/google-research/bert}, for the sake of convenient descriptions.}. The details of these models can be found in~\cite{DBLP:journals/corr/abs-1908-08962}.
We indicate the configurations used for individual phases with the model's suffix.
For example, BERT-QE-\textbf{LLS} indicates that a fine-tuned BERT-\textbf{L}arge is used in phases one and two, and in phase three a fine-tuned BERT-\textbf{S}mall is used.

\begin{table}[!t]
  \label{tab:bert_sizes}
  \centering\resizebox{.4\textwidth}{!}{
  \begin{tabular}{l|l}
  \toprule 
    Size & Configuration \\ \hline
    Tiny (T) & $L=2$, $H=128$, $A=2$ \\
    Small (S) & $L=4$, $H=256$, $A=4$ \\
    Medium (M) & $L=8$, $H=512$, $A=8$ \\
    Base (B) & $L=12$, $H=768$, $A=12$ \\
    Large (L) & $L=24$, $H=1024$, $A=16$ \\
    \toprule
  \end{tabular}}
  \caption{Configurations of different BERT variants.}\label{tab.bertconfigure}
\end{table}

\subsection{Implementation of BERT-QE}\label{sec:config}
Individual documents are decomposed into overlapped passages with 100 words using a sliding window, wherein the stride is 50. 
 For the proposed BERT-QE, in phase two, $k_d=10$ top-ranked documents from the search results of phase one are used, from which $k_c=10$ chunks are selected for expansion, and
 chunk length $m=10$ is used.
In phase one and phase three, the BERT model is used to re-rank the top-1000 documents.
In Section~\ref{sec.analysis}, 
 we also examine the use of different $k_c$ and $m$, namely, 
 $k_c=[5,10,20]$ and $m=[5, 10, 20]$, investigating the impacts of different configurations.

\subsection{Training}

To feed individual query-document pairs into the model, the query $q$ and the document\footnote{As described in Section~\ref{sec.overview}, we actually use the most relevant passage.} $d$ for training
are concatenated and the maximum sequence length is set to 384.
We train BERT using cross-entropy loss for 2 epochs with a batch size of 32 on a TPU v3.
The Adam optimizer~\cite{DBLP:journals/corr/KingmaB14} is used with the learning rate schedule from~\cite{DBLP:journals/corr/abs-1901-04085} with
an initial learning rate of 1e-6. 
We conduct a standard five-fold cross-validation. Namely,
queries are split into five equal-sized partitions. The query partition on Robust04 follows the settings from~\cite{DBLP:conf/sigir/DaiC19}. On GOV2, queries are partitioned by the order of TREC query id in a round-robin manner. In each fold, three partitions are used for training, one is for validation, and the remaining one is for testing. In each fold, we tune the hyper-parameters 
on the validation set and report the performance on test set based on the configurations with the highest NDCG@20 on the validation set\footnote{Results on validation sets can be found in Appendix.}. The ultimate performance is the average among all folds. 

\subsection{Computation of FLOPs}\label{sec.flops}

Akin to literature~\cite{DBLP:conf/acl/LiuZWZDJ20}, we report FLOPs (floating point operations) which measures the computational complexity of models. Similar to~\cite{DBLP:conf/sigir/KhattabZ20}, we report FLOPs that includes all computations in the three phases of BERT-QE.

\section{Results}\label{sec.result}

\begin{table*}[t!]
    \centering
    \resizebox{0.5\textwidth}{!}{\begin{tabular}{l|lll}
    \toprule
      Model & P@20 & NDCG@20 & MAP@1K \\ \hline
     SNRM with PRF & 0.3948 & 0.4391  & 0.2971  \\ 
    NPRF & 0.4064 & 0.4576 &  0.2904 \\
    CEDR  & 0.4667 & 0.5381  & -  \\ 
    Birch(MS$\rightarrow$MB) & 0.4669 & 0.5325  & 0.3691  \\ 
    BERT-Large & $0.4769^{*}$& 0.5397&0.3743\\
    \hline
     BERT-QE-LLL  & $\textbf{0.4888}^{***}$ & $\textbf{0.5533}^{***}$ & $\textbf{0.3865}^{***}$ \\
    \toprule
    \end{tabular}}
    \caption{Compare the effectiveness of BERT-QE-LLL with neural IR models and neural PRF model
    on Robust04 when using title queries. 
    Statistical significance relative to
    Birch(MS$\rightarrow$MB)~\cite{DBLP:conf/emnlp/YilmazYZL19}
    at $\texttt{p-value} < 0.01$, $0.05$, and $0.1$ are denoted as $***$, $**$, and $*$, respectively.}
    \label{tab.neuralir}
\end{table*}

In this section, we report results for the proposed BERT-QE model and compare them to the baseline models.
First, in Section~\ref{sec.mainresult}, we use BERT-Large models for all three phases of BERT-QE.
In Section~\ref{sec.smallerbert}, we evaluate the impact of using smaller BERT models (Table~\ref{tab.bertconfigure}) for the second and third phases
in order to improve the efficiency of the proposed model.

\subsection{Results for BERT-QE-LLL}\label{sec.mainresult}
In this section, we examine the performance of the proposed BERT-QE by comparing it with a range of unsupervised ranking models, 
neural IR models, and re-ranking models based on BERT-Base and BERT-Large. 
We aim at advancing the state-of-the-art ranking performance of BERT-Large, and start with using
BERT-Large for all three phases in BERT-QE.
We denote this variant as BERT-QE-LLL, where the suffix LLL indicates the use of the same fine-tuned BERT-Large in all three phases\footnote{Empirically, the BERT trained on MS MARCO is directly used in phase two, which performs on par with using the fine-tuned BERT according to pilot experiments.}. 

\textbf{The effectiveness of BERT-QE-LLL.} To put our results in context, we first compare BERT-QE-LLL with the reported effectiveness for different neural IR models from literature. Due to the fact that results for GOV2 have not been reported in these works, only the comparisons
on Robust04 are included in Table~\ref{tab.neuralir}.
In comparison with the state-of-the-art results of a fine-tuned BERT-Large, namely, Birch(MS$\rightarrow$MB)~\cite{DBLP:conf/emnlp/YilmazYZL19}, it can be seen that the fine-tuned BERT-Large in this work achieves comparable results.
In addition, BERT-QE-LLL significantly outperforms 
Birch(MS$\rightarrow$MB) at the 0.01 level. The significance tests relative to other models are omitted because 
their result rankings are not available.

As summarized in Table~\ref{tab.main_results}, we further compare BERT-QE-LLL with BERT-Base and BERT-Large on both Robust04 and GOV2.
We also include several unsupervised baselines for reference. 
As can be seen, BERT-Large significantly outperforms all non-BERT baselines by a big margin, regardless of whether query expansion is used.
Thus, only significance tests
relative to BERT-Large 
are shown.
From Table~\ref{tab.main_results}, on Robust04, 
in comparison with BERT-Large,
BERT-QE-LLL could significantly improve the search results on both shallow and
deep pool at 0.01 significant level, achieving a 2.5\% improvement in terms of
NDCG@20 and a 3.3\% improvement for MAP@1K. 
On GOV2, we have similar observations that BERT-QE-LLL could significantly improve BERT-Large on all reported metrics. 

\textbf{The efficiency of BERT-QE.}
Beyond the effectiveness, 
we are also interested in the efficiency of BERT-QE-LLL, for which the FLOPs is reported. The FLOPs per query for BERT-Large is 232.6T, meanwhile BERT-QE-LLL is 2603T.
This means BERT-QE-LLL requires 11.19x more computations than BERT-Large. This is mostly due to the use of BERT-Large models for all three phases as described in Section~\ref{sec.method}.
Note that, 
one may be able to
reduce the time consumption during inference by parallelizing the individual phases of BERT-QE.  
In the following, the efficiency of a model is reported in terms of its relative comparison to BERT-Large, namely, 
in the form of the times of BERT-Large's computational cost.

\begin{table*}[t!]
    \centering\resizebox{\textwidth}{!}{
    \begin{tabular}{l|llll|llll}
    \toprule
  \multirow{2}{*}{Model} & \multicolumn{4}{c}{Robust04} & \multicolumn{4}{c}{GOV2} \\ 
\cline{2-9}
  & P@20   & NDCG@20 & MAP@100    & MAP@1K  & P@20 & NDCG@20 & MAP@100    & MAP@1K \\ \hline
DPH &  0.3616 & 0.4220 & 0.2150  & 0.2512 &  0.5295 & 0.4760  & 0.1731  & 0.3012  \\ \hline
BM25+RM3  & 0.3821  & 0.4407  &  0.2451  & 0.2903 &  0.5634  & 0.4851  &  0.2022  & 0.3350   \\ 
QL+RM3  & 0.3723  & 0.4269  &  0.2314  & 0.2747   & 0.5359  & 0.4568  & 0.1837   & 0.3143  \\
DPH+KL  & 0.3924 & 0.4397  & 0.2528  & 0.3046  & 0.5896 & 0.5122  & 0.2182  & 0.3605  \\ \hline
BERT-Base  &  $0.4653$ & $0.5278$ & $0.3153$ & $0.3652$ & $0.6591$ & $0.5851$ & $0.2535$ & $0.3971$  \\
 BERT-Large & $0.4769$ & $0.5397$ & $0.3238$ & $0.3743$ & $0.6638$ & $0.5932$ & $0.2612$ & $0.4082$  \\
\hline 
BERT-QE-LLL & $\textbf{0.4888}^{***}$ & $\textbf{0.5533}^{***}$ & $\textbf{0.3363}^{***}$ & $\textbf{0.3865}^{***}$ & $\textbf{0.6748}^{***}$ & $\textbf{0.6037}^{***}$ & $\textbf{0.2681}^{***}$ & $\textbf{0.4143}^{***}$ \\ 
\toprule
\end{tabular}}
\caption{Effectiveness of BERT-QE-LLL. Statistical significance relative to BERT-Large at $\texttt{p-value} < 0.01$, $0.05$, and $0.1$ are denoted as $***$, $**$, and $*$, respectively.
}
\label{tab.main_results}
\end{table*}

\begin{table*}
    \centering\resizebox{\textwidth}{!}{
    \begin{tabular}{l|l|llll|llll}
    \toprule
\multirow{2}{*}{Model} &\multirow{2}{*}{FLOPs}& \multicolumn{4}{c}{Robust04} & \multicolumn{4}{c}{GOV2} \\ 
\cline{3-10}
 && P@20   & NDCG@20 & MAP@100    & MAP@1K  & P@20 & NDCG@20 & MAP@100    & MAP@1K \\ \hline
 BERT-Base & 0.28x &  $0.4653$ & $0.5278$ & $0.3153$ & $0.3652$ & $0.6591$ & $0.5851$ & $0.2535$ & $0.3971$  \\
 BERT-Large &1.00x& $0.4769$ & $0.5397$ & $0.3238$ & $0.3743$ & $0.6638$ & $0.5932$ & $0.2612$ & $0.4082$  \\
BERT-QE-LLL  &11.19x& $0.4888^{***}$ & $0.5533^{***}$ & $0.3363^{***}$ & $0.3865^{***}$ & $\textbf{0.6748}^{***}$ & $\textbf{0.6037}^{***}$ & $0.2681^{***}$ & $\textbf{0.4143}^{***}$ \\ 

\hline

BERT-QE-L\textbf{T}L & 11.00x & $0.4855^{***}$ & $0.5500^{***}$ & $0.3318^{***}$ & $0.3821^{***}$ & $0.6691^{**}$ & $0.5986^{*}$ & $0.2663^{***}$ & $0.4138^{***}$ \\
BERT-QE-L\textbf{S}L & 11.00x & $0.4861^{***}$ & $0.5504^{***}$ & $0.3325^{***}$ & $0.3828^{***}$ & $0.6732^{***}$ & $0.6011^{**}$ & $\textbf{0.2685}^{***}$ & $0.4142^{***}$ \\
 BERT-QE-L\textbf{M}L & 11.01x & $\textbf{0.4932}^{***}$ & $\textbf{0.5592}^{***}$ & $\textbf{0.3368}^{***}$ & $\textbf{0.3870}^{***}$ & $0.6715^{**}$ & $0.6013^{*}$ & $0.2675^{*}$ & $0.4136^{*}$ \\
BERT-QE-L\textbf{B}L & 11.05x & $0.4839^{**}$ & $0.5503^{***}$ & $0.3339^{***}$ & $0.3843^{***}$ & $0.6725^{**}$ & $0.6004$ & $0.2639$ & $0.4103$ \\

\hline

BERT-QE-LM\textbf{T}  & \underline{1.03x}& $0.4839^{***}$ & $0.5483^{***}$ & $0.3276^{*}$ & $0.3765$ & $0.6698^{**}$ & $0.5994^{**}$ & $0.2642$ & $0.4098$ \\
BERT-QE-LM\textbf{S} &1.12x&  $0.4910^{***}$ & $0.5563^{***}$ & $0.3315^{***}$ & $0.3810^{**}$ & $0.6658$ & $0.5945$ & $0.2654^{***}$ & $0.4115^{***}$ \\
BERT-QE-LM\textbf{M}  &1.85x& $0.4888^{***}$ & $0.5569^{***}$ & $0.3335^{***}$ & $0.3829^{***}$ & $0.6732^{***}$ & $0.6002^{*}$ & $0.2668^{***}$ & $0.4131^{***}$ \\
BERT-QE-LM\textbf{B}  &3.83x& $0.4906^{***}$ & $0.5580^{***}$ & $0.3367^{***}$ & $0.3858^{***}$ & $0.6728^{***}$ & $0.6011^{**}$ & $0.2649$ & $0.4128^{**}$ \\
\hline

BERT-QE-LL\textbf{T} & 1.20x & $0.4841^{***}$ & $0.5466^{**}$ & $0.3287^{**}$ & $0.3771$ & $0.6695^{**}$ & $0.6009^{**}$ & $0.2650^{**}$ & $0.4110^{*}$ \\
BERT-QE-LL\textbf{S} & \underline{1.30x} & $0.4869^{***}$ & $0.5501^{**}$ & $0.3304^{**}$ & $0.3798^{*}$ & $0.6688^{*}$ & $0.5998^{**}$ & $0.2657^{***}$ & $0.4115^{***}$ \\
BERT-QE-LL\textbf{M} & 2.03x & $0.4811$ & $0.5470$ & $0.3320^{**}$ & $0.3815^{**}$ & $0.6728^{***}$ & $0.6013^{***}$ & $0.2651^{**}$ & $0.4107$ \\
BERT-QE-LL\textbf{B} & 4.01x & $0.4865^{***}$ & $0.5507^{***}$ & $0.3337^{***}$ & $0.3834^{***}$ & $0.6678$ & $0.5984$ & $0.2665^{**}$ & $0.4127^{**}$ \\

\toprule
\end{tabular}}
\caption{Employ different BERT variants for phase two and three in BERT-QE, wherein BERT-Tiny (T), BERT-Small (S), BERT-Medium (M), and BERT-Base (B) are used. Statistical significance relative to BERT-Large at $\texttt{p-value} < 0.01$, $0.05$, and $0.1$ are denoted as $***$, $**$, and $*$, respectively.
}
\label{tab.ablation_final_reranker}
\end{table*}

\subsection{Employing Smaller BERT Variants in BERT-QE}\label{sec.smallerbert}
According to Section~\ref{sec.mainresult}, although with competitive effectiveness, BERT-QE-LLL is very expensive for computation due to the use of BERT-Large in all three phases. In this section, we further explore whether it is possible to 
replace the BERT-Large components with smaller BERT variants from
Table~\ref{tab.bertconfigure}
in the second and third phases, in order to 
further improve the efficiency of the proposed BERT-QE model.
Given that our goal is to improve on BERT-Large, in this work,
we always start with BERT-Large for the first-round ranking.

\textbf{Smaller BERT variants for chunk selector.} As described in Section~\ref{sec.chunkselect}, in the second phase, a BERT model is used to
select text chunks of a fixed length (i.e., $m=10$) by evaluating individual text chunks from the top-$k_d$ documents and selecting the most relevant chunks using a BERT model. Intuitively, compared with ranking a document, evaluating the relevance of a short piece of text is a relatively simple task.
Thus, we examine the use of smaller BERT variants as summarized in the second section (namely, BERT-QE-LXL, where X is T, S, M, or B) in  Table~
\ref{tab.ablation_final_reranker}. 
As shown, compared with using BERT-Large in phase two,
on Robust04, all four BERT-QE variants can outperform BERT-Large significantly at the 0.01 level. Furthermore, BERT-QE-LML can even achieve slightly higher results than BERT-QE-LLL.
On GOV2, on the other hand, 
the uses of BERT-Tiny, BERT-Small, and BERT-Medium could still 
outperform BERT-Large significantly at the 0.05 or 0.1 level,
but with decreasing metrics in most cases.
Overall, for phase two, BERT-Large is a good choice but the smaller BERT variants are also viable. The uses of BERT-Tiny, BERT-Small, and BERT-Medium in phase two can outperform BERT-Large significantly with lower FLOPs.

\textbf{Smaller BERT variants for final re-ranker.}
According to Section~\ref{sec.finalrerank}, 
phase three is the most expensive phase, because a BERT model must compare each document to multiple expansion chunks.
Thus, we further explore the possibility of replacing BERT-Large with smaller BERT variants for phase three.
Based on the results in the previous section, we consider both BERT-Large and BERT-Medium as the chunk selector,
due to the 
superior effectiveness of BERT-QE-LML.
The results are summarized in the third and fourth sections (namely, BERT-QE-LMX and BERT-QE-LLX, where X is T, S, M, or B) of Table~\ref{tab.ablation_final_reranker}.
On Robust04, the use of smaller BERT variants always leads to decreasing effectiveness. However, when using BERT-Small and BERT-Base for the final re-ranking, the corresponding BERT-QE variants always outperform BERT-Large significantly at the 0.1 level. 
BERT-QE-LMM, BERT-QE-LMB, and BERT-QE-LLB can even consistently outperform BERT-Large on all four metrics at the 0.01 level. 
On GOV2, on the other hand, the use of BERT-QE-LMT and BERT-QE-LLM significantly outperforms
BERT-Large on shallow metrics, while BERT-QE-LMS and BERT-QE-LLB outperform BERT-Large on deep metrics.
In addition, BERT-QE-LMM/LLT/LLS consistently outperform BERT-Large on all metrics at 0.1 level.
Overall, 
considering shallow metrics on both datasets, BERT-QE-LMT can outperform
BERT-Large consistently and significantly at the 0.05 level while requiring only 3\% more FLOPs.
On both shallow and deep metrics, BERT-QE-LLS significantly outperforms BERT-Large with 30\% more FLOPs.

\section{Analysis}\label{sec.analysis}
\begin{table*}[ht!]
    \centering\resizebox{0.55\textwidth}{!}{
    \begin{tabular}{l|llll|llll}
    \toprule
  Model& P@20   & NDCG@20   & MAP@1K  \\ \hline
 BERT-Large & $0.4769$ & $0.5397$ & $0.3743$  \\
\hline 
BERT-QE-LLL  & $\textbf{0.4888}^{***}$ & $\textbf{0.5533}^{***}$  & $\textbf{0.3865}^{***}$ \\  \hline
Remove $rel(q, d)$ & $0.4769$ & $0.5372$ & $0.3767$  \\
Chunks from DPH+KL & $0.4759$ & $0.5391$ & $0.3766$  \\
\toprule
\end{tabular}}
\caption{Ablation analyzes for the first-round re-ranker in BERT-QE-LLL, by removing the $rel(q, d)$ from Equation~(\ref{eq.combine}) and by replacing the chunks with the ones selected from top-ranked documents of DPH+KL when computing $rel(q, \mathcal{C})$ in Equation~(\ref{eq.feedbackscore}).
Statistical significance relative to BERT-Large at $\texttt{p-value} < 0.01$, $0.05$, and $0.1$ are denoted as $***$, $**$, and $*$, respectively. 
}
\label{tab.abalation_firstround}
\end{table*}
\subsection{First-round Re-ranker Ablation Analyses}
Intuitively, there are two functions of the first-round ranker:
providing the $rel(q, d)$ score in Equation~(\ref{eq.combine}) used in the final re-ranking, and 
providing the top-$k_d$ documents from which the candidate chunks are selected, which are used to compute $rel(\mathcal{C}, d)$ in Equation~(\ref{eq.feedbackscore}).
In this section, we investigate the impact of the first-round re-ranker from these two perspectives.
In particular, we conduct two ablation analyses: (1) we remove the $rel(q, d)$ from BERT-Large in Equation~(\ref{eq.combine}), but we continue to use the top documents from BERT-Large to select the top-$k_c$ chunks; and (2) we keep the $rel(q, d)$ from BERT-Large in Equation~(\ref{eq.combine}), but we select the top-$k_c$ chunks from documents returned by the unsupervised DPH+KL model.
The results are summarized in Table~\ref{tab.abalation_firstround}.
For the first ablation, when $rel(q, d)$ from BERT-Large is not used, BERT-QE cannot outperform BERT-Large.
Similarly, in the second ablation, selecting chunks from the documents returned by DPH+KL also prevents BERT-QE from outperforming BERT-Large.
These results highlight the importance of both functions of the first-round re-ranker. That is, we need a powerful 
model for the first-round re-ranker to provide ranking score $rel(q, d)$ and the high-quality feedback documents for the chunk selector.

\subsection{Hyper-parameter study}

\begin{figure}[!t]
    \centering
    \includegraphics[width=0.48\textwidth]{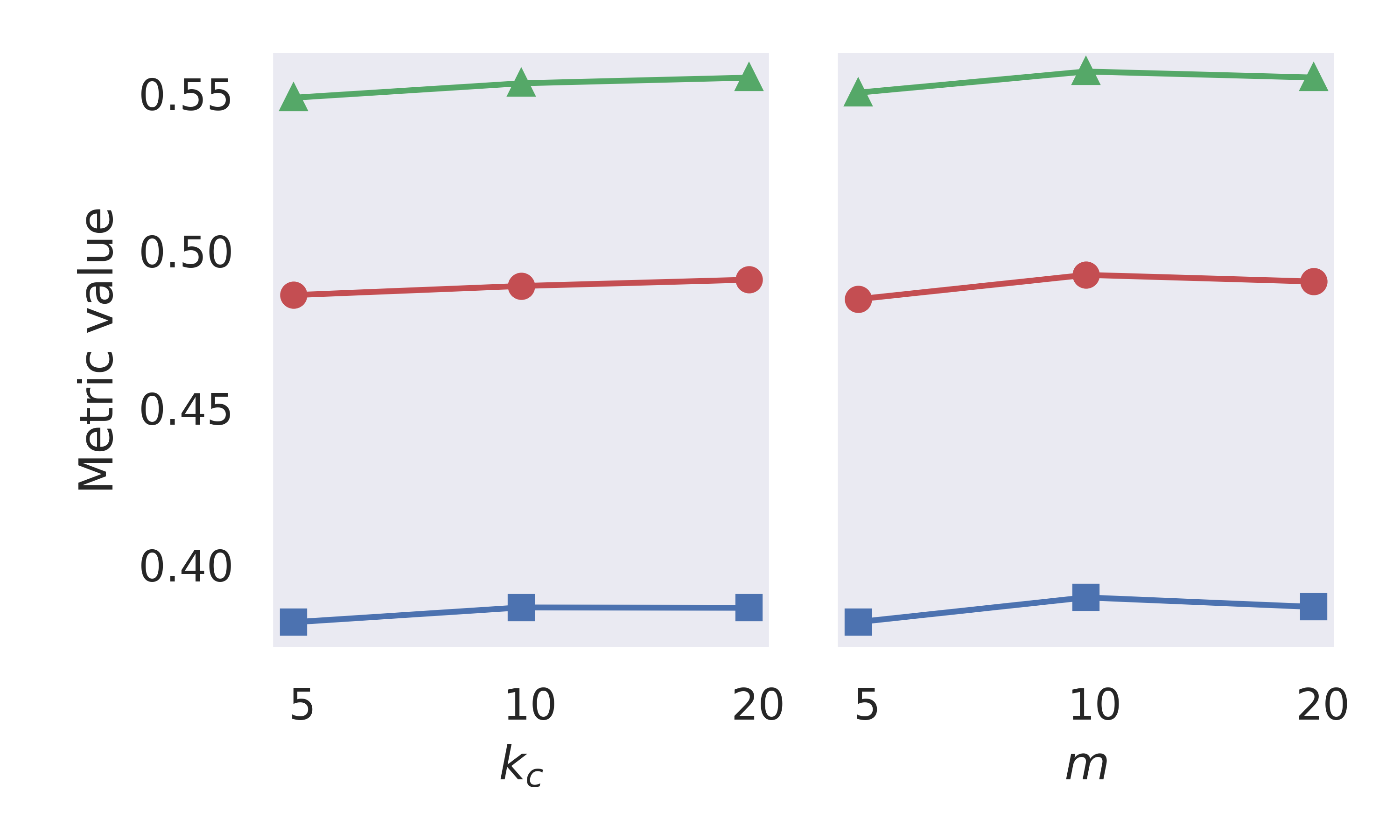}
    \caption{Performance of BERT-QE with different configurations of $k_c$ and $m$. The $\circ$, $\bigtriangleup$, $\Box$ correspond to results in terms of P@20, NDCG@20, and MAP@1K, respectively.}
    \label{fig:hyper-params}
\end{figure}

There are two hyper-parameters in the proposed BERT-QE, namely $k_c$ and $m$. $k_c$ is the number of chunks used in the final-round re-ranking as described in Equation~(\ref{eq.feedbackscore}). 
Meanwhile, 
the chunk size $m$  balances between contextual information and noise. Results for different hyper-parameter settings on Robust04 are shown in Figure~\ref{fig:hyper-params}. 
For $k_c$,
it can be seen that $k_c=10, 20$ achieve similar
performance, while $k_c=5$ reduces the results.  As the computational cost of phase three is proportional to $k_c$ and the performance gaps between $k_c=10$ and $k_c=20$ are actually quite small, $k_c=10$ is a reasonable and robust configuration.
Among different settings of $m$,
$m=10$ achieves the best performance and therefore is used in the proposed model.

\section{Related Work}\label{sec.related}
 
 \noindent \textbf{BERT for IR.} 
 Inspired by the success of contextualized models like BERT on NLP tasks, 
 ~\citet{DBLP:journals/corr/abs-1901-04085} examine the performance of BERT on the passage re-ranking tasks using MS MARCO and TREC-CAR datasets, and demonstrate superior performances compared with the existing shallow ranking models like
 Co-PACRR~\cite{DBLP:conf/wsdm/HuiYBM18} and KNRM~\cite{DBLP:conf/sigir/XiongDCLP17}.
 Thereafter,
 the application of contextualized BERT model in ranking tasks have attracted many attentions.
 ~\citet{DBLP:conf/sigir/DaiC19} split a document into fixed length passages and use a BERT ranker to predict the relevance of each passage independently. The score of the first passage, the best passage, or the sum of all passage scores is used as the document score. 
 ~\citet{DBLP:conf/sigir/MacAvaneyYCG19} incorporate BERT's classification vector into existing neural models,
 including DRMM~\cite{DBLP:conf/cikm/GuoFAC16}, PACRR~\cite{DBLP:conf/emnlp/HuiYBM17}, and KNRM~\cite{DBLP:conf/sigir/XiongDCLP17},
 demonstrating promising performance boosts.
 ~\citet{DBLP:conf/emnlp/YilmazYZL19} transfer models across different domains and aggregate sentence-level evidences to rank documents. 
~\citet{DBLP:journals/corr/abs-1910-14424} propose a multi-stage ranking architecture with BERT that can trade off quality against latency.
 ~\citet{DBLP:conf/www/WuMLZZZM20} propose the context-aware Passage-level Cumulative Gain to aggregate passage relevance representations scores, which is incorporated into a BERT-based model for document ranking.
In addition to these efforts, 
this work further proposes to exploit the contextualized BERT model to expand the original queries in the proposed BERT-QE framework, boosting the ranking performance by using the pseudo feedback information
 effectively.

 \noindent \textbf{Query expansion} has long been applied to make use of the pseudo relevance feedback information~\cite{DBLP:conf/ictir/HuiHLW11} to
 tackle the vocabulary mismatch problem. Keyword query expansion methods, such as Rocchio's algorithm~\cite{rocchio1971relevance} and
 the KL query expansion model~\cite{DBLP:phd/ethos/Amati03},
 have been shown to be effective when applied to text retrieval tasks. Moreover, ~\citet{DBLP:conf/sigir/MetzlerC07} propose to expand beyond unigram keywords by using a Markov random field model.
 Some query expansion methods use word embeddings to find the relevant terms to the query~\cite{DBLP:conf/acl/0001MC16,DBLP:conf/ictir/ZamaniC16a}. 
  ~\citet{DBLP:conf/sigir/CaoNGR08} perform query expansion by using classification models to select expansion terms. 
  NPRF~\cite{DBLP:conf/emnlp/LiSHWHYSX18} incorporates existing neural ranking models like DRMM~\cite{DBLP:conf/cikm/GuoFAC16} into an end-to-end neural PRF framework.
Rather than expanding the query, ~\citet{DBLP:journals/corr/abs-1904-08375} propose a document expansion method named Doc2query, which uses a neural machine translation method to generate queries that each document might answer. Doc2query is further improved by docTTTTTquery~\cite{Nogueira2019doc} which replaces the seq2seq transformer with T5~\cite{DBLP:journals/corr/abs-1910-10683}. 
 ~\citet{DBLP:conf/sigir/MacAvaneyN0TGF20b} construct query and passage representations and perform passage expansion based on term importance.
 Despite the promising results of the above document expansion methods for passage retrieval, they are so far only applied to short text retrieval tasks to avoid excessive memory consumption.
 In comparison with these established expansion models, the proposed BERT-QE aims at better selecting and incorporating the information pieces from feedback, by taking advantages of the BERT model in
 identifying relevant information.

\section{Conclusion}\label{sec.conclusion}
This work proposes a novel expansion model, coined as BERT-QE, 
to better select relevant information
for query expansion.
Evaluation on the Robust04 and GOV2 test collections 
confirms that  
BERT-QE significantly outperforms BERT-Large with 
relatively small extra computational cost (up to 30\%).
In future work, we plan to further improve the efficiency of BERT-QE, by combining the proposed BERT-QE with the pre-computation techniques proposed recently~\cite{DBLP:conf/sigir/KhattabZ20,DBLP:conf/sigir/MacAvaneyN0TGF20}, wherein most of the computations could be performed offline.

\section*{Acknowledgments}
This research work is supported by the National Natural Science Foundation of China under Grants no. U1936207 and 61772505, Beijing Academy of Artiﬁcial Intelligence (BAAI2019QN0502), the Youth Innovation Promotion Association CAS (2018141), and University of Chinese Academy of Sciences.

\bibliographystyle{acl_natbib}
\bibliography{anthology,emnlp2020}

\begin{thebibliography}{38}
\expandafter\ifx\csname natexlab\endcsname\relax\def\natexlab#1{#1}\fi

\bibitem[{Amati(2003)}]{DBLP:phd/ethos/Amati03}
Giambattista Amati. 2003.
\newblock \emph{Probability models for information retrieval based on
  divergence from randomness}.
\newblock Ph.D. thesis, University of Glasgow, {UK}.

\bibitem[{Amati et~al.(2007)Amati, Ambrosi, Bianchi, Gaibisso, and
  Gambosi}]{DBLP:conf/trec/AmatiABGG07}
Gianni Amati, Edgardo Ambrosi, Marco Bianchi, Carlo Gaibisso, and Giorgio
  Gambosi. 2007.
\newblock Fub, {IASI-CNR} and university of tor vergata at {TREC} 2007 blog
  track.
\newblock In \emph{{TREC}}, volume Special Publication 500-274. National
  Institute of Standards and Technology {(NIST)}.

\bibitem[{Bajaj et~al.(2018)Bajaj, Campos, Craswell, Deng, Gao, Liu, Majumder,
  McNamara, Mitra, Nguyen, Rosenberg, Song, Stoica, Tiwary, and
  Wang}]{bajaj2018ms}
Payal Bajaj, Daniel Campos, Nick Craswell, Li~Deng, Jianfeng Gao, Xiaodong Liu,
  Rangan Majumder, Andrew McNamara, Bhaskar Mitra, Tri Nguyen, Mir Rosenberg,
  Xia Song, Alina Stoica, Saurabh Tiwary, and Tong Wang. 2018.
\newblock Ms marco: A human generated machine reading comprehension dataset.
\newblock \emph{CoRR}, abs/1611.09268v3.

\bibitem[{Cao et~al.(2008)Cao, Nie, Gao, and
  Robertson}]{DBLP:conf/sigir/CaoNGR08}
Guihong Cao, Jian{-}Yun Nie, Jianfeng Gao, and Stephen Robertson. 2008.
\newblock Selecting good expansion terms for pseudo-relevance feedback.
\newblock In \emph{{SIGIR}}, pages 243--250. {ACM}.

\bibitem[{Clarke et~al.(2004)Clarke, Craswell, and
  Soboroff}]{DBLP:conf/trec/ClarkeCS04}
Charles L.~A. Clarke, Nick Craswell, and Ian Soboroff. 2004.
\newblock Overview of the {TREC} 2004 terabyte track.
\newblock In \emph{{TREC}}, volume Special Publication 500-261. National
  Institute of Standards and Technology {(NIST)}.

\bibitem[{Dai and Callan(2019)}]{DBLP:conf/sigir/DaiC19}
Zhuyun Dai and Jamie Callan. 2019.
\newblock Deeper text understanding for {IR} with contextual neural language
  modeling.
\newblock In \emph{{SIGIR}}, pages 985--988. {ACM}.

\bibitem[{Devlin et~al.(2019)Devlin, Chang, Lee, and
  Toutanova}]{DBLP:conf/naacl/DevlinCLT19}
Jacob Devlin, Ming{-}Wei Chang, Kenton Lee, and Kristina Toutanova. 2019.
\newblock {BERT:} pre-training of deep bidirectional transformers for language
  understanding.
\newblock In \emph{{NAACL-HLT} {(1)}}, pages 4171--4186. Association for
  Computational Linguistics.

\bibitem[{Diaz et~al.(2016)Diaz, Mitra, and Craswell}]{DBLP:conf/acl/0001MC16}
Fernando Diaz, Bhaskar Mitra, and Nick Craswell. 2016.
\newblock Query expansion with locally-trained word embeddings.
\newblock In \emph{{ACL} {(1)}}. The Association for Computer Linguistics.

\bibitem[{Guo et~al.(2016)Guo, Fan, Ai, and Croft}]{DBLP:conf/cikm/GuoFAC16}
Jiafeng Guo, Yixing Fan, Qingyao Ai, and W.~Bruce Croft. 2016.
\newblock A deep relevance matching model for ad-hoc retrieval.
\newblock In \emph{{CIKM}}, pages 55--64. {ACM}.

\bibitem[{Hui et~al.(2011)Hui, He, Luo, and Wang}]{DBLP:conf/ictir/HuiHLW11}
Kai Hui, Ben He, Tiejian Luo, and Bin Wang. 2011.
\newblock A comparative study of pseudo relevance feedback for ad-hoc
  retrieval.
\newblock In \emph{{ICTIR}}, volume 6931 of \emph{Lecture Notes in Computer
  Science}, pages 318--322. Springer.

\bibitem[{Hui et~al.(2017)Hui, Yates, Berberich, and
  de~Melo}]{DBLP:conf/emnlp/HuiYBM17}
Kai Hui, Andrew Yates, Klaus Berberich, and Gerard de~Melo. 2017.
\newblock {PACRR:} {A} position-aware neural {IR} model for relevance matching.
\newblock In \emph{{EMNLP}}, pages 1049--1058. Association for Computational
  Linguistics.

\bibitem[{Hui et~al.(2018)Hui, Yates, Berberich, and
  de~Melo}]{DBLP:conf/wsdm/HuiYBM18}
Kai Hui, Andrew Yates, Klaus Berberich, and Gerard de~Melo. 2018.
\newblock Co-pacrr: {A} context-aware neural {IR} model for ad-hoc retrieval.
\newblock In \emph{{WSDM}}, pages 279--287. {ACM}.

\bibitem[{Khattab and Zaharia(2020)}]{DBLP:conf/sigir/KhattabZ20}
Omar Khattab and Matei Zaharia. 2020.
\newblock Colbert: Efficient and effective passage search via contextualized
  late interaction over {BERT}.
\newblock In \emph{{SIGIR}}, pages 39--48. {ACM}.

\bibitem[{Kingma and Ba(2015)}]{DBLP:journals/corr/KingmaB14}
Diederik~P. Kingma and Jimmy Ba. 2015.
\newblock Adam: {A} method for stochastic optimization.
\newblock In \emph{{ICLR}}.

\bibitem[{Lavrenko and Croft(2001)}]{DBLP:conf/sigir/LavrenkoC01}
Victor Lavrenko and W.~Bruce Croft. 2001.
\newblock Relevance-based language models.
\newblock In \emph{{SIGIR}}, pages 120--127. {ACM}.

\bibitem[{Li et~al.(2018)Li, Sun, He, Wang, Hui, Yates, Sun, and
  Xu}]{DBLP:conf/emnlp/LiSHWHYSX18}
Canjia Li, Yingfei Sun, Ben He, Le~Wang, Kai Hui, Andrew Yates, Le~Sun, and
  Jungang Xu. 2018.
\newblock {NPRF:} {A} neural pseudo relevance feedback framework for ad-hoc
  information retrieval.
\newblock In \emph{{EMNLP}}, pages 4482--4491. Association for Computational
  Linguistics.

\bibitem[{Lin et~al.(2016)Lin, Crane, Trotman, Callan, Chattopadhyaya, Foley,
  Ingersoll, MacDonald, and Vigna}]{DBLP:conf/ecir/LinCTCCFIMV16}
Jimmy~J. Lin, Matt Crane, Andrew Trotman, Jamie Callan, Ishan Chattopadhyaya,
  John Foley, Grant Ingersoll, Craig MacDonald, and Sebastiano Vigna. 2016.
\newblock Toward reproducible baselines: The open-source {IR} reproducibility
  challenge.
\newblock In \emph{{ECIR}}, volume 9626 of \emph{Lecture Notes in Computer
  Science}, pages 408--420. Springer.

\bibitem[{Liu et~al.(2020)Liu, Zhou, Wang, Zhao, Deng, and
  Ju}]{DBLP:conf/acl/LiuZWZDJ20}
Weijie Liu, Peng Zhou, Zhiruo Wang, Zhe Zhao, Haotang Deng, and Qi~Ju. 2020.
\newblock Fastbert: a self-distilling {BERT} with adaptive inference time.
\newblock In \emph{{ACL}}, pages 6035--6044. Association for Computational
  Linguistics.

\bibitem[{MacAvaney et~al.(2020{\natexlab{a}})MacAvaney, Nardini, Perego,
  Tonellotto, Goharian, and Frieder}]{DBLP:conf/sigir/MacAvaneyN0TGF20}
Sean MacAvaney, Franco~Maria Nardini, Raffaele Perego, Nicola Tonellotto, Nazli
  Goharian, and Ophir Frieder. 2020{\natexlab{a}}.
\newblock Efficient document re-ranking for transformers by precomputing term
  representations.
\newblock In \emph{{SIGIR}}, pages 49--58. {ACM}.

\bibitem[{MacAvaney et~al.(2020{\natexlab{b}})MacAvaney, Nardini, Perego,
  Tonellotto, Goharian, and Frieder}]{DBLP:conf/sigir/MacAvaneyN0TGF20b}
Sean MacAvaney, Franco~Maria Nardini, Raffaele Perego, Nicola Tonellotto, Nazli
  Goharian, and Ophir Frieder. 2020{\natexlab{b}}.
\newblock Expansion via prediction of importance with contextualization.
\newblock In \emph{{SIGIR}}, pages 1573--1576. {ACM}.

\bibitem[{MacAvaney et~al.(2019)MacAvaney, Yates, Cohan, and
  Goharian}]{DBLP:conf/sigir/MacAvaneyYCG19}
Sean MacAvaney, Andrew Yates, Arman Cohan, and Nazli Goharian. 2019.
\newblock {CEDR:} contextualized embeddings for document ranking.
\newblock In \emph{{SIGIR}}, pages 1101--1104. {ACM}.

\bibitem[{Macdonald et~al.(2012)Macdonald, McCreadie, Santos, and
  Ounis}]{macdonald2012puppy}
Craig Macdonald, Richard McCreadie, Rodrygo~LT Santos, and Iadh Ounis. 2012.
\newblock From puppy to maturity: Experiences in developing terrier.
\newblock \emph{Proc. of OSIR at SIGIR}, pages 60--63.

\bibitem[{Metzler and Croft(2007)}]{DBLP:conf/sigir/MetzlerC07}
Donald Metzler and W.~Bruce Croft. 2007.
\newblock Latent concept expansion using markov random fields.
\newblock In \emph{{SIGIR}}, pages 311--318. {ACM}.

\bibitem[{Nogueira and Cho(2019)}]{DBLP:journals/corr/abs-1901-04085}
Rodrigo Nogueira and Kyunghyun Cho. 2019.
\newblock Passage re-ranking with {BERT}.
\newblock \emph{CoRR}, abs/1901.04085.

\bibitem[{Nogueira et~al.(2020)Nogueira, Jiang, and
  Lin}]{DBLP:journals/corr/abs-2003-06713}
Rodrigo Nogueira, Zhiying Jiang, and Jimmy Lin. 2020.
\newblock Document ranking with a pretrained sequence-to-sequence model.
\newblock \emph{CoRR}, abs/2003.06713.

\bibitem[{Nogueira and Lin(2019)}]{Nogueira2019doc}
Rodrigo Nogueira and Jimmy Lin. 2019.
\newblock From doc2query to doc{TTTTT}query.
\newblock \emph{Technical report}.

\bibitem[{Nogueira et~al.(2019{\natexlab{a}})Nogueira, Yang, Cho, and
  Lin}]{DBLP:journals/corr/abs-1910-14424}
Rodrigo Nogueira, Wei Yang, Kyunghyun Cho, and Jimmy Lin. 2019{\natexlab{a}}.
\newblock Multi-stage document ranking with {BERT}.
\newblock \emph{CoRR}, abs/1910.14424.

\bibitem[{Nogueira et~al.(2019{\natexlab{b}})Nogueira, Yang, Lin, and
  Cho}]{DBLP:journals/corr/abs-1904-08375}
Rodrigo Nogueira, Wei Yang, Jimmy Lin, and Kyunghyun Cho. 2019{\natexlab{b}}.
\newblock Document expansion by query prediction.
\newblock \emph{CoRR}, abs/1904.08375.

\bibitem[{Qiao et~al.(2019)Qiao, Xiong, Liu, and
  Liu}]{DBLP:journals/corr/abs-1904-07531}
Yifan Qiao, Chenyan Xiong, Zheng{-}Hao Liu, and Zhiyuan Liu. 2019.
\newblock Understanding the behaviors of {BERT} in ranking.
\newblock \emph{CoRR}, abs/1904.07531.

\bibitem[{Raffel et~al.(2019)Raffel, Shazeer, Roberts, Lee, Narang, Matena,
  Zhou, Li, and Liu}]{DBLP:journals/corr/abs-1910-10683}
Colin Raffel, Noam Shazeer, Adam Roberts, Katherine Lee, Sharan Narang, Michael
  Matena, Yanqi Zhou, Wei Li, and Peter~J. Liu. 2019.
\newblock Exploring the limits of transfer learning with a unified text-to-text
  transformer.
\newblock \emph{CoRR}, abs/1910.10683.

\bibitem[{Rocchio(1971)}]{rocchio1971relevance}
J.~Rocchio. 1971.
\newblock Relevance feedback in information retrieval.
\newblock In Gerard Salton, editor, \emph{The {SMART} retrieval system:
  experiments in automatic document processing}, pages 313--323. Prentice Hall,
  Englewood, Cliffs, New Jersey.

\bibitem[{Turc et~al.(2019)Turc, Chang, Lee, and
  Toutanova}]{DBLP:journals/corr/abs-1908-08962}
Iulia Turc, Ming{-}Wei Chang, Kenton Lee, and Kristina Toutanova. 2019.
\newblock Well-read students learn better: The impact of student initialization
  on knowledge distillation.
\newblock \emph{CoRR}, abs/1908.08962.

\bibitem[{Voorhees(2004)}]{DBLP:conf/trec/Voorhees04b}
Ellen~M. Voorhees. 2004.
\newblock Overview of the {TREC} 2004 robust track.
\newblock In \emph{{TREC}}, volume Special Publication 500-261. National
  Institute of Standards and Technology {(NIST)}.

\bibitem[{Wu et~al.(2020)Wu, Mao, Liu, Zhan, Zheng, Zhang, and
  Ma}]{DBLP:conf/www/WuMLZZZM20}
Zhijing Wu, Jiaxin Mao, Yiqun Liu, Jingtao Zhan, Yukun Zheng, Min Zhang, and
  Shaoping Ma. 2020.
\newblock Leveraging passage-level cumulative gain for document ranking.
\newblock In \emph{{WWW} '20: The Web Conference 2020, Taipei, Taiwan, April
  20-24, 2020}, pages 2421--2431. {ACM} / {IW3C2}.

\bibitem[{Xiong et~al.(2017)Xiong, Dai, Callan, Liu, and
  Power}]{DBLP:conf/sigir/XiongDCLP17}
Chenyan Xiong, Zhuyun Dai, Jamie Callan, Zhiyuan Liu, and Russell Power. 2017.
\newblock End-to-end neural ad-hoc ranking with kernel pooling.
\newblock In \emph{{SIGIR}}, pages 55--64. {ACM}.

\bibitem[{Yilmaz et~al.(2019)Yilmaz, Yang, Zhang, and
  Lin}]{DBLP:conf/emnlp/YilmazYZL19}
Zeynep~Akkalyoncu Yilmaz, Wei Yang, Haotian Zhang, and Jimmy Lin. 2019.
\newblock Cross-domain modeling of sentence-level evidence for document
  retrieval.
\newblock In \emph{{EMNLP/IJCNLP} {(1)}}, pages 3488--3494. Association for
  Computational Linguistics.

\bibitem[{Zamani and Croft(2016)}]{DBLP:conf/ictir/ZamaniC16a}
Hamed Zamani and W.~Bruce Croft. 2016.
\newblock Embedding-based query language models.
\newblock In \emph{{ICTIR}}, pages 147--156. {ACM}.

\bibitem[{Zamani et~al.(2018)Zamani, Dehghani, Croft, Learned{-}Miller, and
  Kamps}]{DBLP:conf/cikm/ZamaniDCLK18}
Hamed Zamani, Mostafa Dehghani, W.~Bruce Croft, Erik~G. Learned{-}Miller, and
  Jaap Kamps. 2018.
\newblock From neural re-ranking to neural ranking: Learning a sparse
  representation for inverted indexing.
\newblock In \emph{{CIKM}}, pages 497--506. {ACM}.

\end{thebibliography}

\newpage
\appendix

\section{Appendices}
\label{sec:appendix}

\subsection{Interpolation Parameters in BERT-QE}

\begin{table}[h!]
    \centering\resizebox{0.48\textwidth}{!}{
    \begin{tabular}{c|cccccc}
    \toprule
  & \multicolumn{6}{c}{Robust04} \\ 
\cline{2-7}
 Fold & P@20   & NDCG@20 & MAP@100    & MAP@1K & $\alpha$ & $\beta$ \\ \hline
 1 & 0.4730 &  0.5606 & 0.3247 & 0.3765 &  0.4 & 0.9  \\
 2 & 0.4900 &  0.5666 & 0.3909 & 0.4362 &  0.4 & 0.8  \\
 3 & 0.4740 &  0.5328 & 0.2941 & 0.3471 &  0.4 & 0.9  \\
 4 & 0.4684 &  0.5213 & 0.2940 & 0.3440 &  0.6 & 0.9   \\
 5 & 0.5400 &  0.5868 & 0.3709 & 0.4233 &  0.3 & 0.9   \\
 \hline \hline
 & \multicolumn{6}{c}{GOV2} \\ \hline
  1 & 0.6233 & 0.5728 & 0.2257 & 0.3621 & 0.4 & 0.9 \\
 2 & 0.7397 & 0.6675 & 0.3046 & 0.4334 & 0.7 & 0.9 \\
 3 & 0.7167 & 0.6177 & 0.2558 & 0.4456 & 0.1 & 0.7 \\
 4 & 0.6850 & 0.6027 &  0.2718 &  0.4140 & 0.4 & 0.8  \\
 5 & 0.6300 & 0.5731 & 0.2860 & 0.4240 & 0.4 & 0.8  \\
 
 \toprule
\end{tabular}}
\caption{Results on \textit{validation sets}, as well as the chosen interpolation parameters $\alpha$ and $\beta$ based on \textit{validation sets} for BERT-QE-LLL. 
}
\label{tab.supplement}
\end{table}

There are two hyper-parameters in BERT-QE, namely $\alpha$ and $\beta$, both of which are interpolation coefficients. $\alpha$ is introduced in 
Equation~(\ref{eq.combine}). 
In addition,
akin to~\cite{DBLP:conf/emnlp/YilmazYZL19},
there is an interpolation with the initial ranking, i.e., DPH+KL, 
which has been applied to all models, including BERT-QE and baselines,
where $\beta$ is the hyper-parameter.
As shown in the following equation,
$M(q,d)$ denotes the scores from a re-ranking model, e.g., BERT-QE model.
$I(q,d)$ denotes the scores from the initial ranking, namely, DPH+KL.
$\alpha$ and $\beta$ are both tuned on the validation set through grid search on (0,1) with stride 0.1. 
The models with best nDCG@20 on validation sets are chosen.
Different configurations of $\alpha$ and $\beta$ and the corresponding results are summarized in Table~\ref{tab.supplement}.

\begin{equation*}
    \mathit{final\_score} = \beta \cdot \log(M(q,d))
    + (1-\beta) \cdot I(q,d)
\end{equation*}

\begin{table}[h!]
    \centering\resizebox{0.3\textwidth}{!}{
    \begin{tabular}{l|c}
  \toprule 
    Size & \# of parameters \\ \hline
    Tiny (T) & 4M \\
    Small (S) & 11M \\
    Medium (M) & 41M \\
    Base (B) & 109M \\
    Large (L) & 335M \\
    \toprule
  \end{tabular}}
\caption{Number of parameters in BERT variants.}
\label{tab.num_params}
\end{table}

\subsection{Number of parameters in BERT variants}

We list the number of parameters in different BERT variants
used in BERT-QE in Table~\ref{tab.num_params}.

\end{document}